\documentclass[twocolumn,nofootinbib,
showpacs,prl,aps,floatfix]{revtex4}

\usepackage{graphicx}

\newcommand {\fabs}[1] {\left| #1 \right|}
\newcommand {\fabsq}[1] {\left| #1 \right|^2}
\newcommand {\fexp} [1] {\exp \left( #1 \right)}

\renewcommand{\Re}{\mbox{Re}}

\begin{document}

\newcommand{\GHz}{\,\mbox{GHz}}
\newcommand{\MHz}{\,\mbox{MHz}}
\newcommand{\m}{\,\mbox{m}}
\newcommand{\mus}{\,\mu\mbox{s}}

\title{Simultaneous arrival of information in absorbing wave guides}

\author {A. Ruschhaupt}

\author {J. G. Muga}
\affiliation{Departamento de Qu\'{\i}mica-F\'{\i}sica,
UPV-EHU,\\
Apartado 644, 48080 Bilbao, Spain}

\begin{abstract}
We demonstrate 
that the temporal peak generated 
by specific electromagnetic pulses 
may arrive at different positions simultaneously in an absorbing
wave guide. The effect can 
be used for triggering several devices all at once at unknown
distances from
the sender or generally to transmit information 
so that it arrives at the same time to receivers at different,
unknown locations.
This simultaneity cannot be realized by 
the standard transmission methods. 
\end{abstract}
\maketitle

In a previous paper \cite{delgado.2004} we have described a
surprising effect, namely, that the
temporal peak of a quantum wave from a source
with a sharp onset 
may arrive at different locations
simultaneously in absorbing media.
This ``ubiquitous peak'' (UP) 
effect was  
demonstrated for the Schr\"odiger equation,
where, unlike the Hartman effect,  
it holds at arbitrarily large distances
(see \cite{delgado.2004} for further differences
with the Hartman effect);  
and also 
for a relativistic
wave equation, but limited to distances where 
the time of arrival of the peak is larger than the 
time of the very first front
\cite{delgado.2004}.  
In this letter we demonstrate that the effect is also found
within a broad 
spatial range  
in an absorbing 
wave guide when the source emits (more realistic) smoothed pulses
instead of a perfectly sharp 
step signal.
The optimal carrier frequency is barely below the cut-off but, 
at variance with 
other ``ultrafast'' wave phenomena based on anomalous dispersion
in absorbing media \cite{garrett.1970,chu.1982,TAT2001,RS02}, which  
depend on the dominance of the carrier (central) 
frequency associated with faster than light,    
infinite 
or negative group velocities, the ubiquitous peak is, at each  
position, 
dominated by the saddle-point contributions above the
cut-off frequency. It is thus a fundamentally
different phenomenon.

One advantage of the wave guide with respect to the quantum particle 
described by the Schr\"odinger equation is that the effect may be 
implemented more easily and could be
measured in a non-invasive way \cite{gersen.2003}. 
It also makes possible to trigger 
several devices at the same time 
or transmit information so that it arrives simultaneously 
at unknown locations.  
This cannot be achieved by standard  
transmission methods because   
a receiver could not resend an information bit to the 
closest one 
faster than the velocity of light in vacuum $c$;
nor can we design the timing of 
a series of signals 
from the source so that they arrive simultaneously at different receivers 
if their locations are unknown.

We assume a wave guide in $z$-direction filled with a
homogeneous, isotropic, dielectric, dispersive, absorbing medium, i.e.
we end up with the following wave equation 
(see e.g. \cite{jackson.book} for details)
\begin{eqnarray}
\left[\frac{\partial^2}{\partial z^2} +
\frac{1}{c^2} \, \left(i \frac{\partial}{\partial t}\right)^2
\eta^2\!\!\left(i \frac{\partial}{\partial t}\right)
- \gamma^2 \right] \phi (z,t) = 0,
\label{eq_main}
\end{eqnarray}
where $\eta(\omega)$ is the complex refraction index
of the medium,
and $\gamma$ is the mode eigenvalue of the waveguide.

We assume that the refraction index is given by 
the Lorentz model, 
\begin{eqnarray*}
\eta(\omega)
= \sqrt{1 - \omega_p^2 \frac{\omega^2 - \omega_L^2 - 2i \delta \omega}
{(\omega^2-\omega_L^2)^2 + 4 \delta^2 \omega^2}}, 
\end{eqnarray*}
and consider, for a frequency interval around $\omega_0$, 
$|\omega-\omega_0|<\Delta$,  the conditions  
$\delta \gg \fabs{(\omega^2-\omega_L^2)/(2\omega)}$ and
$\delta \gg \omega_p^2/(2\omega)$, 
so that 
\begin{eqnarray}
\eta(\omega) \approx \sqrt{1 + i \frac{\omega_p^2}{2\delta \omega}}
\approx 1 + i \frac{\omega_p^2}{4\delta\omega}
\equiv 1 + i \frac{1}{\omega} n_1.
\label{refr}
\end{eqnarray}
(With this choice Eq. (\ref{eq_main}) is similar to
the Klein-Gordon equation discussed in \cite{delgado.2004}.)

\begin{figure}
\begin{center}
\includegraphics[angle=-90,width=\linewidth]{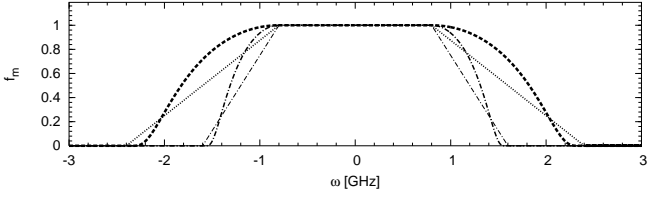}
\end{center}
\caption[]{Window functions $f_m$;
$\Delta\omega = 0.8 \GHz$;
$m=2, \alpha=0.8 \GHz$
(thin dashed-dotted line),
$m=2, \alpha=1.6 \GHz$
(thin dotted line),
$m=3, \alpha=0.8 \GHz$
(thick dashed-dotted line),
$m=3, \alpha=1.6 \GHz$
(thick dotted line).}
\label{fig1}
\end{figure}

We also assume ``source'' boundary conditions with 
the value of $\phi (0,t) \equiv \phi_0 (t)$ given for all $t$, 
and require $\phi (z,t) = 0$ for $z>0$, $t<0$,  which
fixes a unique solution $\phi (z,t)$.
The UP effect was found first for the 
sharp-onset source function $e^{-i\omega_0 t} \Theta(t)$
but we shall now examine smoother variants,  
\begin{eqnarray}
\phi_{0m} (t) & = &
\frac{i}{2\pi} \int_{-\infty}^{\infty} d\omega f_m(\omega - \omega_0)
\frac{e^{-i\omega t}}{\omega - \omega_0 + i0}.
\label{eq_init}
\end{eqnarray}
In the reference case $f_m (\omega) = f_1 (\omega) \equiv 1$,  
$\phi_{01} (t) = e^{-i\omega_0 t} \Theta(t)$.
The solution of Eq. (\ref{eq_main}) with the boundary condition 
in Eq. (\ref{eq_init}) fulfilling the demand
that $\phi_m (z,t) = 0$ for $z>0$, $t<0$ is
\begin{eqnarray}
\phi_m (z,t) =
\frac{i}{2\pi} \int_{-\infty}^{\infty}
d\omega f_m(\omega - \omega_0)
\frac{e^{i k(\omega) z - i \omega t}}{\omega - \omega_0 + i 0},
\label{eq_solution}
\end{eqnarray}
with the dispersion relation
\begin{eqnarray*}
k(\omega) = \sqrt{\frac{\omega^2}{c^2} \eta^2(\omega) - \gamma^2}
= \frac{1}{c} \sqrt{\left(\omega + i n_1\right)^2 - 
\omega_c^2},
\end{eqnarray*}
and $\omega_c=c\gamma$ being the cut-off frequency.
We may apply the saddle-pole approximation of the integral
in Eq. (\ref{eq_solution})
for the case $f_m(\omega) = 1$.
The approximation $\phi \approx \Theta(g) \phi_p + \phi_{s+} + \phi_{s-}$
can be found by following the steps described e.g. in \cite{buettiker.1998}. 
The saddle points are 
$\omega_{s\pm} = \pm \beta - i n_1$
with $\beta = \omega_c/\sqrt{1-z^2/(c^2 t^2)}$.
The pole contribution,  
if 
%
$0 < g(z,t) = (\omega_0 - \beta)(\omega_0 \beta - \omega_c^2)
 + n_1\sqrt{(\beta^2 - \omega_c^2)
(-\omega_0^2 + 2 \beta \omega_0 - \omega_c^2)}$, 
%
is
$\phi_p (z,t) = - \fexp{i z k(\omega_0) - i \omega_0 t}$, whereas 
the saddle contributions are
\begin{eqnarray*}
\phi_{s\pm} (z,t) &=& \frac{i}{\sqrt{2\pi}} \frac{z \sqrt{\mp i}
\sqrt{\omega_c}}
{c t^{3/2} \sqrt[4]{1 - \frac{z^2}{c^2t^2}}}\\
&\times& \frac{\fexp{-t n_1 \mp i t \omega_c
\sqrt{1-\frac{z^2}{c^2 t^2}}}}{\omega_c \mp \sqrt{1-\frac{z^2}{c^2 t^2}}
\left(\omega_0 + i n_1\right)}.
\end{eqnarray*}
If there is no pole contribution, 
$\phi \approx \phi_{s+} + \phi_{s-}$, 
one may easily obtain lower and upper envelops for the oscillating signal, 
$I_- \le \fabsq{\phi_{s+} + \phi_{s-}} \le I_+$.  
Let us now examine the exact solutions for other ``window functions''
$f_m$ with a central plateau (see Fig. \ref{fig1} for examples), 
\begin{eqnarray*}
f_2 (\omega) &=& \left\{\begin{array}{lcl}
1 &\!:\!& 0 \le \fabs{\omega} < \Delta\omega\\
1 - \frac{\fabs{\omega}-\Delta\omega}{\alpha} 
  &\!:\!& \Delta \omega \le \fabs{\omega} < \Delta\omega + \alpha\\
0 &\!:\!& \Delta \omega + \alpha  < \fabs{\omega}
\end{array}\right.\\
f_3 (\omega) &=& \left\{\begin{array}{lcl}
1 &\!:\!& 0 \le \fabs{\omega} < \Delta\omega\\
\fexp{-\frac{(\fabs{\omega}-\Delta\omega)^2}{\alpha^2-(\fabs{\omega}-\Delta\omega)^2}}
  &\!:\!& \Delta \omega \le \fabs{\omega} < \Delta\omega + \alpha\\
0 &\!:\!& \Delta \omega + \alpha  < \fabs{\omega}
\end{array}\right..
\end{eqnarray*}
%
%
Since $f_2$ and $f_3$ are non-zero only in a range
around $0$ it is enough that the form of the refraction index, 
Eq. (\ref{refr}), 
is fulfilled in an interval
$2(\Delta w+\alpha)$ around $\omega_0$.

\begin{figure}
\begin{center}
\includegraphics[angle=-90,width=\linewidth]{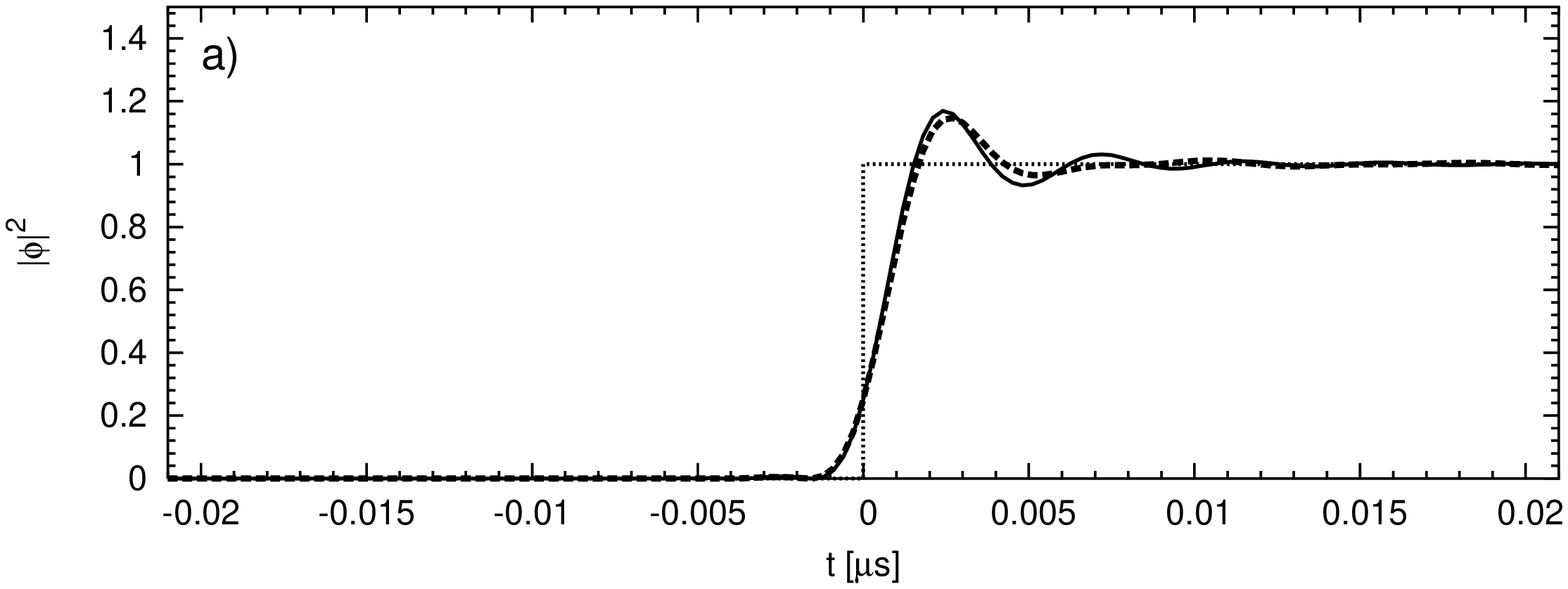}

\includegraphics[angle=-90,width=\linewidth]{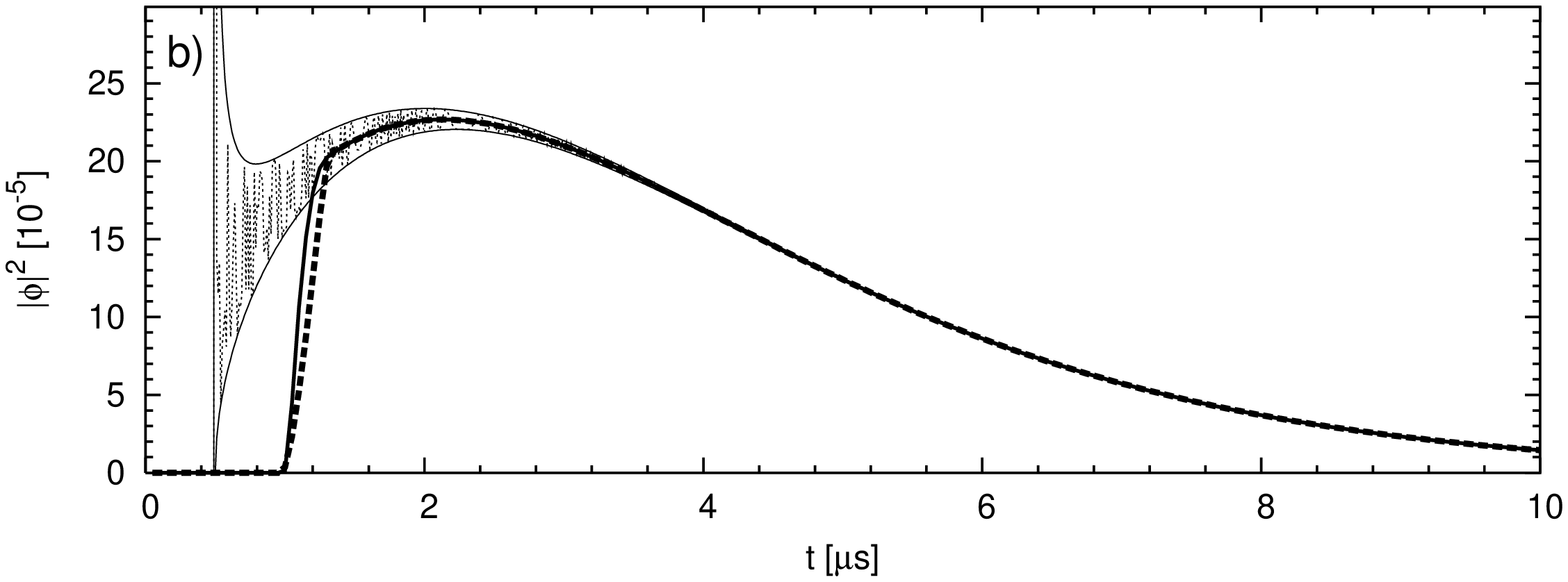}
\end{center}
\caption[]{Wave function $\fabsq{\phi_m (z,t)}$ versus $t$ for 
$\omega_0 = 9.49 \GHz$, $\omega_c = 9.5 \GHz$, $\Delta\omega = 0.8 \GHz$,
$\alpha = 0.8 \GHz$,  $n_1 = 0.2 \MHz$; 
$m=1$ (thin dotted line), $m=2$ (thick dotted line), $m=3$ (thick solid
line).
The thin solid lines are the envelops $I_-$ and $I_+$
; (a) $z=0$, (b) $z=150 \m$.}
\label{fig2}
\end{figure}

Figure \ref{fig2}a shows $\fabsq{\phi_m (0, t)}$ for different $m$ and 
Fig. \ref{fig2}b shows $\fabsq{\phi_m (z,t)}$ for $z = 150 \m$.  
A consequence of the smoothing of the signal onset at the source is 
the cancellation of the oscillations
between the two envelops
, i.e., a much simpler signal structure.
Also, the very first  
sharp causal front for $f_1$ is substituted by a smooth increase
for the window functions $f_{2,3}$, 
but the maximum around $t=2\,\mu$s remains. 
We are interested in the
time $\tau_T$ of this maximum for $m=2,3$. 
%
%
%
\begin{figure}
\begin{center}
\includegraphics[angle=-90,width=\linewidth]{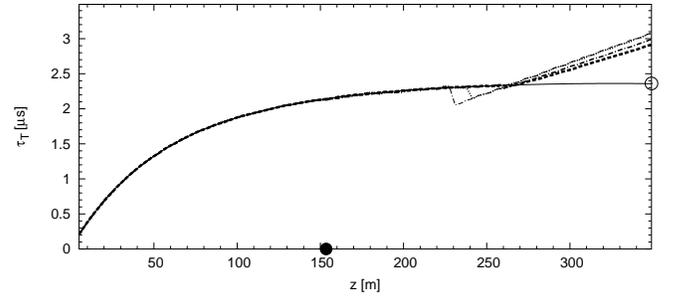}
\end{center}
\caption[]{$\tau_T$ versus $z$
for different source signals; the filled dot
indicates $z_{min}$; the unfilled dot indicates $\tau_M$;
$\omega_c = 9.5 \GHz$, $\omega_0 = 9.49 \GHz$,
$n_1= 0.2 \MHz$, $\Delta\omega = 0.8 \GHz$;
$m=2, \alpha=0.8 \GHz$
(thin dashed-dotted line),
$m=2, \alpha=1.6 \GHz$
(thin dotted line),
$m=3, \alpha=0.8 \GHz$
(thick dashed-dotted line),
$m=3, \alpha=1.6 \GHz$
(thick dotted line);
$\tau_{Ts+}$ (solid line).}
\label{fig3}
\end{figure}
%
The times $\tau_T$  are depicted in Fig. \ref{fig3} versus $z$ for different 
$m$ and $\alpha$: 
$\tau_T$ for $m=2,3$ is nearly independent
of $z$ for intermediate values of $z$, see also Fig. 4, 
whereas at large $z$ the 
maximum behaves ``normally''
and grows linearly with $z$.

Figure \ref{fig3} also shows that the time $\tau_T$
is independent of the type of edge $m=2$ (discontinuous) or $m=3$
(continuous) 
and of the value of $\alpha$
in a wide spatial range. In other words, the effect is  stable 
with respect to the detailed form of the window edges,
an important result for implementing it.  
One may e.g. substitute the 
strict band-limitation imposed by the finite support of $f_{2,3}$
by infinite-support windows.   

The solution can be approximated 
by the contribution of the positive saddle $\phi_{s+}$
up to an upper critical distance, as shown in  
Fig. \ref{fig3}, where  
the solid line is the time of the temporal maximum $\tau_{Ts+}$
calculated from the positive saddle (i.e. by taking $\phi \approx \phi_{s+}$).
Because $\mbox{Re}(\omega_{s+})=\omega_c/\sqrt{1-z^2/(c^2 t^2)} > \omega_c$ in
the region of the UP effect, it follows that
the peak is predominantly composed by frequencies above
the cut-off $\omega_c$.
The group velocity calculated at the positive saddle \cite{tanaka.1986}
is always smaller than $c$, but it is unrelated here to the 
peak's behaviour.

\begin{figure}
\begin{center}
\includegraphics[angle=-90,width=\linewidth]{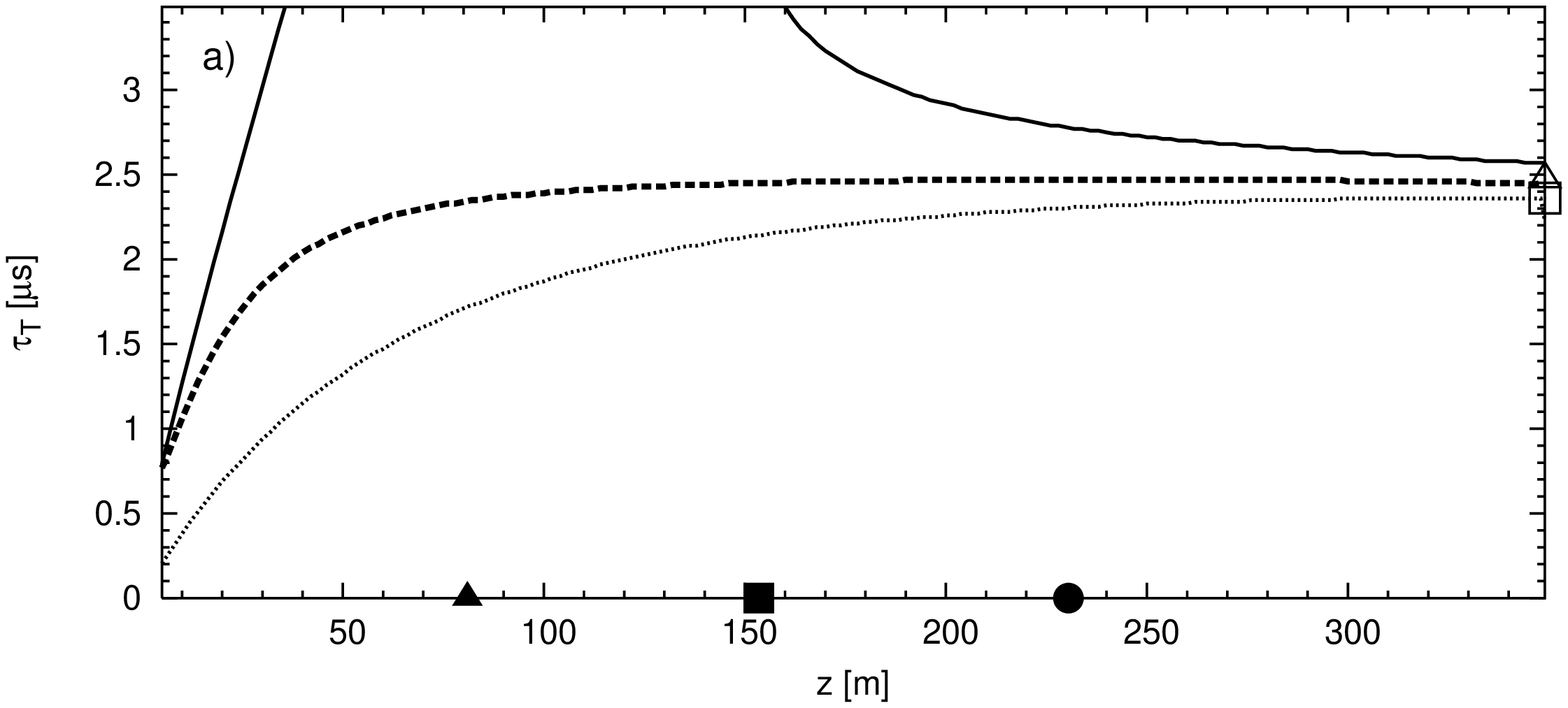}

\includegraphics[angle=-90,width=\linewidth]{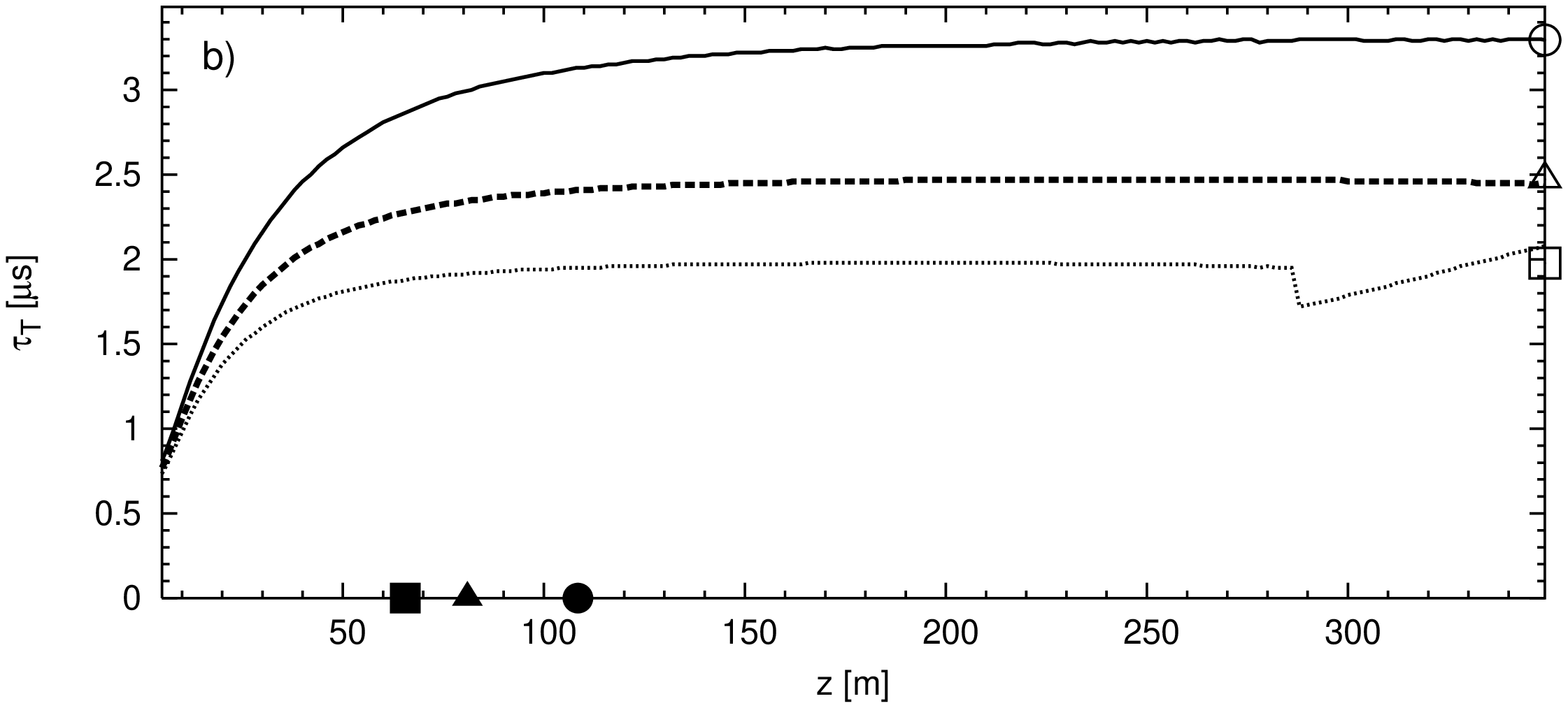}
\end{center}
\caption[]{Time of the temporal maximum $\tau_T$ versus $z$
for the initial function $\phi_{03}$;
$\omega_c = 9.5 \GHz$, $\Delta\omega = 2.0 \GHz$,
$\alpha=0.4 \GHz$; the filled symbols
indicate $z_{min}$;  the unfilled symbols
indicate $\tau_M$;  (a) $n_1= 0.2 \MHz$;
$\omega_0 = 9.49 \GHz$ (thin dotted line/box), 
$\omega_0 = 9.499 \GHz$ (thick dotted line/triangle),
$\omega_0 = 9.51 \GHz$ (thick solid line/circle);
(b) $\omega_0 = 9.499 \GHz$;
$n_1 = 0.25 \MHz$ (thin dotted line/box), 
$n_1 = 0.2 \MHz$ (thick dotted line/triangle),
$n_1 = 0.15 \MHz$ (thick solid line/circle).}
\label{fig4}
\end{figure}

$\tau_{Ts+}(z)$ may be used 
to estimate a lower value for the start of the effect: $z_{min}$ is defined 
as the smallest $z$ where $\fabs{\frac{d}{dz} \tau_{Ts+}(z)} < 1/c$, 
so that the temporal maximum ``moves'' beyond $z_{min}$ faster than
$c$ without violating causality in any way.
Assuming $n_1\ll\omega_0$ and 
$1<\xi\equiv\omega_c/\omega_0<3/2^{3/2}$
we can find an analytical formula for the maximal
value of $\tau_{Ts+}$,
\begin{eqnarray}
\tau_M = \frac{1}{2 n_1} \, \frac{3 - 2 \xi^2
- 3 (\xi^2 - 1)^{2/3}}{\xi^2}\approx \frac{1}{2n_1},
\label{eq_taumax}
\end{eqnarray}
which gives the arrival time of the peak in
the region where the UP effect 
holds. 
The values $\tau_M$ are shown in Fig. \ref{fig3}
and Fig. \ref{fig4} with empty symbols.

The approximation
$\phi\approx\phi_{s+}$ breaks down when the saddle point 
reaches the edge of the window function, i.e.
$\Re(\omega_{s+}) \ge \omega_0 + \Delta\omega$; 
then the UP effect also breaks down and $\tau_T$ grows 
linearly with $z$.
From the condition
$\Re(\omega_{s+})|_{t=\tau_M} = \omega_0 + \Delta\omega$
the upper boundary for the effect is given by  
$z_{max}= c \tau_{M} \sqrt{1-\omega_c^2/
(\omega_0 + \Delta\omega)^2}$, which 
increases with $\Delta\omega$.
This value of $z_{max}$ coincides with the penetration length
of $\omega=\omega_0 + \Delta\omega$ defined by 
$l=1/\{2{\rm{Im}}[k(\omega)]\}$.
However, $z_{max}$ cannot be arbitrarily large 
since the maximum of the saddle eventually  
vanishes. (The value $z_M$ where this occurs is given by a lengthy 
expression. 
In the range of parameters considered in the examples  
$z_{max}<z_M$ so that $z_{max}$ is the true upper bound.)

Let us examine what happens by changing $\omega_0$ or
$n_1$.
According to Fig. \ref{fig4}a, the effect exists also 
for carrier frequencies above the cut-off,
$\omega_0>\omega_c$,  
but in that case a much greater peak appears at small 
$z$ which
travels with finite velocity, and the simultaneous arrival effect
is only seen at larger 
distances from the source when the main forerunner has not yet arrived 
and a much smaller peak is formed first. 
For the applications described below 
(simultaneous triggering or sending information 
that arrives simultaneously 
to different receivers)
it is more convenient to use energies just below the cut-off, because 
the traveling forerunner does not 
exist, the attenuation is minimal, and the spatial range in which 
the effect holds becomes maximal.  
In Fig. \ref{fig4}b we can see that 
a stronger absorption leads to faster arrival of the peak.
The upper and lower limits of the effect diminish 
when $n_1$ is increased,
but detection becomes more difficult because of the attenuation of the 
signal.

%
%

\begin{figure}
\begin{center}
\includegraphics[angle=-90,width=\linewidth]{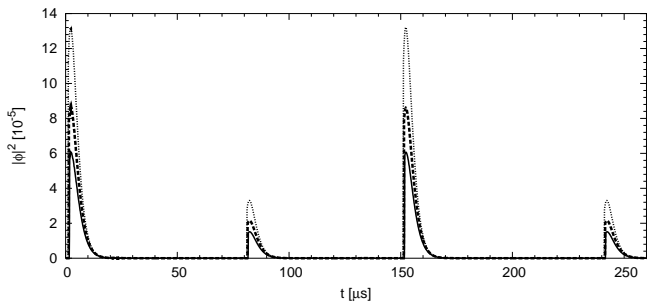}
\end{center}
\caption[]{Wave function $\fabsq{\Phi (z,t)}$ versus $t$ for 
$\omega_0 = 9.49 \GHz$, $\omega_c = 9.5 \GHz$, $\Delta\omega = 2.0 \GHz$,
$\alpha = 0.4 \GHz$,  $n_1 = 0.2 \MHz$;
$b_1=1.0$, $b_2=-0.5$, $t_2=80\mus$,
$b_3=1.0$, $t_3=150\mus$, $b_4=-0.5$, $t_4=240\mus$;
$z=200 \m$ (thin dotted line), $z=250 \m$ (thick dotted line),
$z=300 \m$ (solid line).}
\label{fig5}
\end{figure}

In the following we shall illustrate how this effect
can be used to send a triggering signal or 
information to receivers at unknown distances in such a way
that the information arrives at all receivers at nearly the same time.
We shall use a wave function at the source of the form
\begin{eqnarray*}
\Phi_0 (t) & = & b_1 \phi_{03} (t) 
+ b_2 \phi_{03}(t-t_2)
\\
&+& b_3 \phi_{03}(t-t_3)
+ b_4 \phi_{03}(t-t_4).
\end{eqnarray*}
As Eq. (\ref{eq_main}) is linear, $\Phi (z,t)$
can be found by adding the solutions corresponding to  
each term separately.
An example of $\fabsq{\Phi (z,t)}$ is plotted for different $z$
in Fig.~\ref{fig5} where 
the times of the maxima are nearly independent
of $z$.
%
%
%
\begin{figure}
\begin{center}
\includegraphics[angle=-90,width=\linewidth]{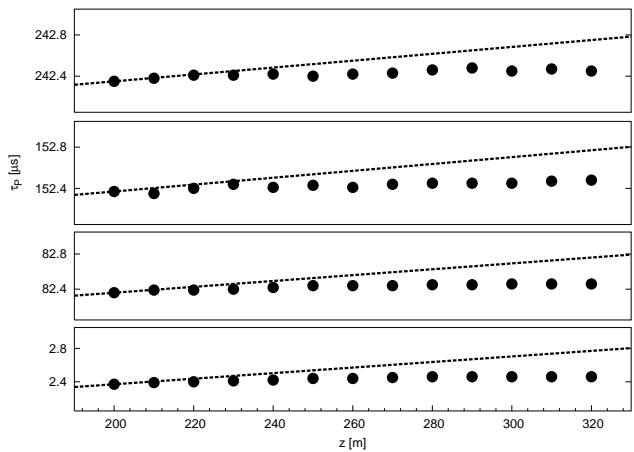}
\end{center}
\caption[]{Times of finding the peaks $\tau_P$ versus $z$ (dots);
$l_0 = 0.5 \times 10^{-5}$, $\Delta t= 0.1 \mus$;
the lines have gradient $1/c$; for the other parameters
see Fig. \ref{fig5} and text for details.}
\label{fig6}
\end{figure}
%
Suppose that the receivers are located at unknown 
distances from the
source and that they do not have synchronized clocks
(only their unit time intervals are equal).
Each receiver gets $\fabsq{\Phi(z,t)}$ and
may use the following operational procedure to find the peaks:
$\tau_P$ is the time of finding a peak (the peak is at time
$\tau_P - \Delta t$) if
$\fabsq{\Phi(z,\tau_P - \Delta t)} > \fabsq{\Phi(z,t)}$
for $\tau_P - \Delta t > t \ge \tau_P$. Moreover there should be
a noise level $l_0$ so that $\fabsq{\Phi(z,t)}$ is assumed as zero
if $\fabsq{\Phi(z,t)} < l_0$. After finding a peak the search for the
next peak is started if $\fabsq{\Phi(z,t)} < l_0$. The noise
level may establish an upper limit $z_{noise}$ for the effect 
more strict than $z_{max}$, beyond which 
the attenuation makes impossible in practice
to distinguish the peak.

The resulting times $\tau_P$ of the different peaks are
plotted in Fig. \ref{fig6}.
Clearly the receivers find the peaks at nearly the same times $\tau_P$.
For comparison,
lines with slope $1/c$ are also plotted.
In all cases the receivers get the peak
earlier than if the nearest receiver sends a light signal to them
when it finds the peak.
This shows that the effect can be used to trigger receivers at different
unknown distances if we use only the first maximum as
the triggering signal.
Moreover it is possible to send bits of information coded in the value
of $\fabsq{\Phi(z,t)}$ at the time of the maxima. The 
height of the first two peaks may be used for calibration, e.g.
the higher peak may represent a logical $1$ and the 
lower peak a logical $0$, whereas 
the following peaks carry the message.
The signal in Fig. \ref{fig5}, for example, 
would represent a sequence $1010$.
It is not necessary that the peaks are sent at equal 
time intervals.  

Summarizing, we have shown that the temporal maxima generated by 
specific wave pulses in an absorbing wave guide 
arrive simultaneously at receivers in a 
broad and partly controllable domain.
The task of sending information to arrive at different receivers 
simultaneously
is different from the question of superluminal
velocities because the information always arrives subluminally 
\cite{wynne.2002,stenner.2003} 
and it will be possible in principle to send information
faster to a single fixed receiver than with the present effect. 

The ubiquitous peak is dominated by above-cut-off frequencies
so it is of a fundamentally different origin from effects based
on superluminal tunneling
and on negative and/or superluminal group velocities
\cite{garrett.1970, chu.1982,TAT2001,RS02, nimtz.1997,
chiao.1997,kuzmich.2001,stenner.2003,buettiker.2003}.
In the case described by the Schr\"odiger equation \cite{delgado.2004} it 
is closer in nature to the over-the-barrier,  
saddle dominated peak that arrives at the B\"uttiker-Landauer time  
in a non-absorbing  waveguide \cite{MB02, VRS02},
and in fact tends to it continuously when 
the absorption vanishes. However, that peak ``moves'' with a semiclassical 
tunneling velocity whereas in the absorbing medium it 
appears everywhere simultaneously within the domain of
the effect.

\begin{acknowledgments}
We acknowledge support by 
UPV-EHU (00039.310-13507/2001), ``Ministerio de
Ciencia y Tecnolog\'\i a''
and FEDER (BFM2003-01003).  
AR acknowledges support by the German Academic Exchange
Service (DAAD). 
\end{acknowledgments}

\end{document}